# Loss-Compensating Si Photonics Signal Routers

**P. F. J**ARSCHEL[*]**, M. C. M. M. S**OUZA**, R. B. M**ERLO **AND N. C. F**RATESCHI

*Device Research Laboratory, Applied Physics Department, "Gleb Wataghin" Physics Institute, University of Campinas - UNICAMP, 13083-859 Campinas-SP, Brazil*
*\*jarschel@ifi.unicamp.br*

**Abstract:** We propose and demonstrate a low-cost integrated photonic chip fabricated in a SOI foundry capable of simultaneously routing and amplifying light in a chip. This device is able to compensate insertion losses in photonic routers. It consists of standard Si/SiO2 ring resonators with Er:Al2O3 as the upper cladding layer, employed using only one simple post-processing step. This resulted in a measured on/off gain of 0.9 dB, with a footprint smaller than 0.002 mm$^2$, and expected bit rates as high as 40Gb/s based on the resonance quality-factor. We show that the on/off gain value can be further increased using coupled rings to reach net gain values of 4 dB.

## 1. Introduction

One of the challenges faced by Silicon Photonics is maintaining the optical power along a complex integrated circuit that has a dense sequence of devices. Mimicking Erbium-Doped Fiber Amplifiers (EDFAs) functionality in the chip level is an attractive approach to help overcoming this problem. To this end, Erbium-Doped Waveguide Amplifiers (EDWAs) using Aluminum Oxide ($Al_2O_3$) as the waveguide core material have been demonstrated [1]. Such amplifiers benefit from high concentration of Er+ ions in the $Al_2O_3$ matrix, up to 100 times higher than possible in $SiO_2$ [2]. However, the process of integrating such alumina waveguides to Silicon Photonic chips departs from CMOS technology and demands complex techniques to achieve reasonable coupling between the Si and $Al_2O_3$ waveguides. Also, this increases chip design time, demands a great deal of additional post processing steps, and reduces yield. A simpler approach consists of using a doped film as the Si waveguide top cladding instead. Since a considerable fraction of the propagating mode in silicon waveguides is outside the core, this method can allow efficient light amplification and even lasing [3].

A Silicon Photonics building block of particular interest is the ring resonator, often used to filter, route, and generally control the flow of signals [4]. However, when using add/drop ports for signal routing, insertion losses are always present [5]. It is possible to minimize this effect by changing the coupling strength between the rings and the feeding/extracting waveguides, though this leads to a decreased Quality Factor (Q) and the consequent loss of filtering resolution. In a complex photonic circuit, several of these components may be cascaded, and the added losses may give rise to information deterioration.

In order to mitigate this insertion loss problem, one may employ EDWA as buffers distributed along the chip. However, a full amplifier may occupy a relatively large area, limiting the room for other components. In this context, we propose and demonstrate a device that acts simultaneously as a filter and amplifier. It consists of silicon ring resonators with erbium-doped alumina cladding, which can be easily fabricated in SOI Photonics Foundries, with minimal post-processing.

## 2. Modeling

To effectively model our proposed device, we have to combine two concepts: the output spectra of a ring resonator with additional add/drop ports, described by the Transfer Matrix method [6], and the rate equations for light amplification [7]. These can be approximately coupled by considering the rate of change in the electrical field amplitude inside the ring ($E_{R1}$ to $E_{R2}$ and $E_{R3}$ to $E_{R4}$, as seen in Fig. 1a) as the sum of the propagation loss and the expected optical gain for a waveguide of the same length (Fig. 1b). To this end, we calculated the overlap factors by FEM simulations, obtaining 0.33 for the signal wavelength (~1550 nm) and 0.3 for the pump wavelength (~1480 nm, resonant with the ring).

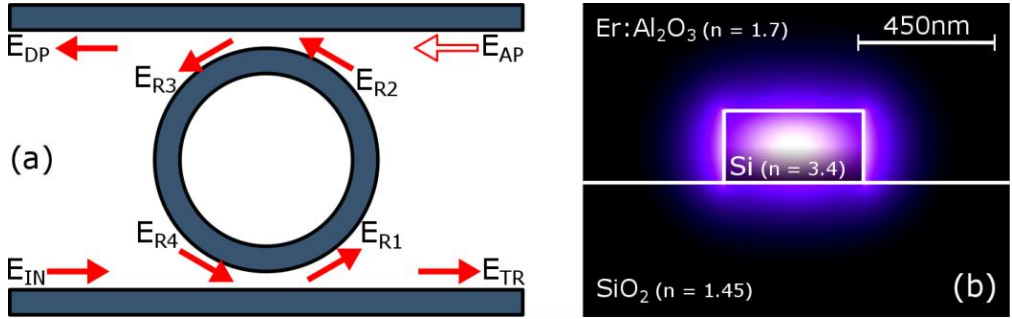

Fig. 1: (a) Schematic of a ring resonator with add/drop ports with the electric fields amplitudes and light propagation direction indicated at specific points, according to the TMM model; (b) FEM simulation of the quasi-TE mode of the resonator waveguide with 20 µm bending radius. The SOI waveguide is clad with Er-doped alumina film (n = 1.7).

The output spectra at the drop port ($E_{DP}$), shown in From Fig. 2a and Fig. 2b, were obtained by using typical values for the Er concentration ($3 \times 10^{20}$ cm$^{-3}$), absorption/emission cross-sections (~0.6 pm$^2$, varies with wavelength) [8], and lifetime (8 ms). The dimensions and materials considered are the same as the devices we fabricated: 20 µm of ring radius, 450 nm x 220 nm waveguide, 200 nm of gap between the ring and bus waveguide, and propagation loss of 2 dB/cm. Finally, an input signal of 1 µW and pump powers up to 10 mW were chosen for these results. We can observe the influence of the doped material in the output spectra, and the difference between the not pumped/pumped curves is as high as 1 dB at the 1536 nm resonance.

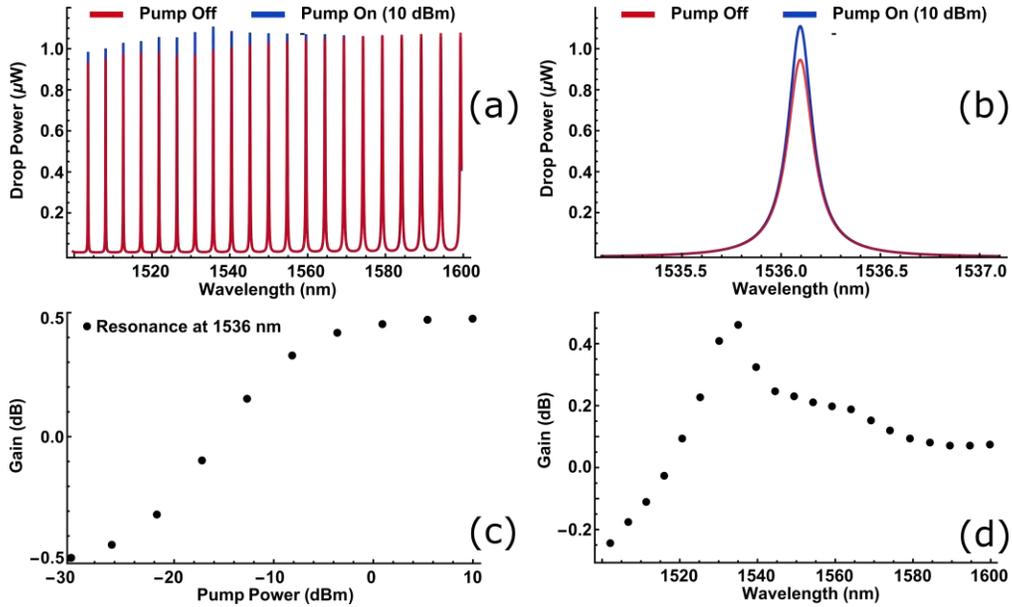

Fig. 2: Simulated device performance. (a) Drop-port spectra for ON (10 mW, blue) and OFF (red) pump conditions (b) Detail of one resonance; (c) Output gain for the 1536.1 nm resonance as a function of pump power; (d) Spectral gain.

By varying the pump power and comparing the output from this resonance with the input, we observe the occurrence of optical gain from -10 dBm of pump and upwards, saturating with 0.47 dB of net gain (Fig. 2c), due to depletion of the excited Er ions. If we now take the saturated gain for each resonance, we see that they form a curve that resembles that of the Erbium emission cross-section, with peak performance at the 1535 nm region (Fig. 2d). This shows that both concepts were successfully coupled on the developing of this model. If we perform the same calculation for an identical device, but with an undoped cladding (Er concentration = 0), we obtain 0.4 dB of loss, showing that our modelled device can not only completely compensate the insertion loss for this design, but also provide a small gain under these conditions.

## 3.  Experimental results and Discussion

To experimentally demonstrate the proposed device, we proceed with the fabrication using unclad chips containing samples of Si Ring Resonators manufactured by a SOI Photonics Foundry (IMEC), via standard CMOS-compatible processes. We deposited the Er-Doped $Al_2O_3$ film on top of the whole chip using the Co-Sputtering technique [9]. To serve as a control sample, we deposited an undoped $Al_2O_3$ film on top of one of the chips, instead of the Er-doped.

The resonators consisted of an external ring (20 µm radius), with two smaller rings (5 µm radius) internally coupled to it, as seen in Fig. 3a and Fig. 3b. The coupled inner rings create additional resonances with increased Q factor, creating the possibility of higher amplification efficiency.

The setup shown in Fig. 3c was used to measure the output spectra. Two tunable lasers are employed: one for the signal, which is coupled to the device with a micro-lensed fiber after passing through a circulator, and another for the pump, which also passes through a circulator before being coupled to the ring from the drop port. Both lasers must have their polarization carefully adjusted for proper coupling with the waveguide; this is done with standard fiber polarization controllers. The signal exiting the device via the drop port is coupled to the lensed fiber (the same as the pump input), circulates to an optical filter to remove any traces of the pump, and is finally measured in a photodetector. The remaining pump exits through the signal input, and is monitored with another photodetector after passing through the circulator. The lasers and photodetectors are computer controlled.

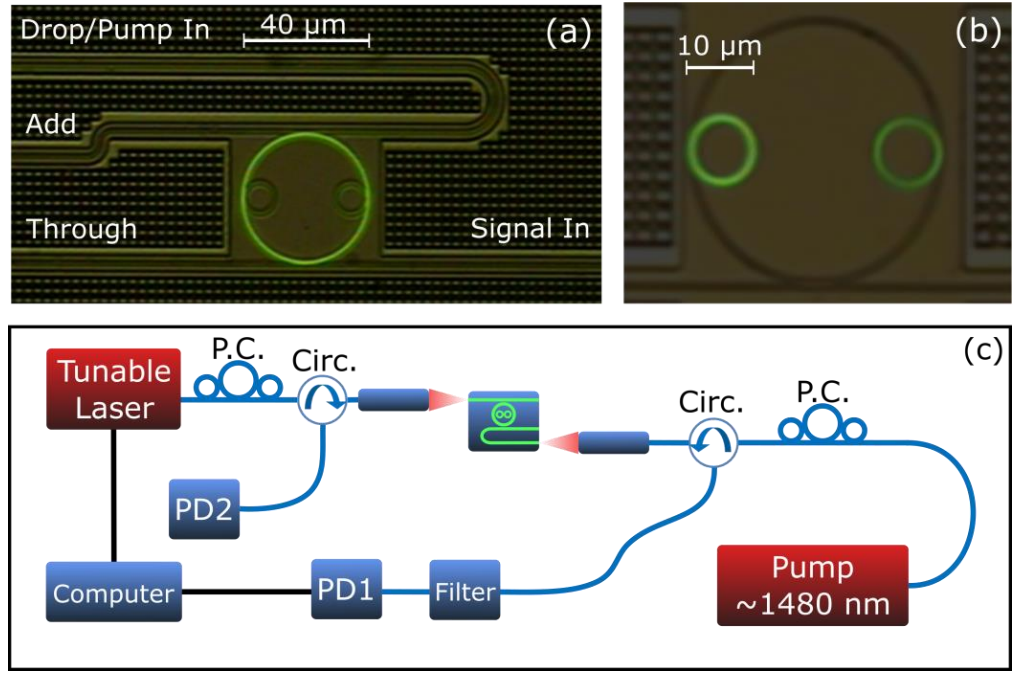

Fig. 3: Fabricated device and indicated ports when the outer ring is pumped (a), and when the inner rings are pumped (b). (c) Schematic of the measurement setup. PD: photodetector; Filter: band-pass optical filter; P.C.: polarization controller; Circ.: circulator.

By setting the pump wavelength to the closest resonance to 1480 nm, sweeping the signal (10 μW) wavelength and measuring the output from the drop ports, we obtained the spectra shown in Fig. 4a and Fig. 4b. There is a clear difference in behavior between the doped and control (undoped) samples. When the doped sample is pumped (10 dBm), there is an increase in output power of 1 dB for the shown resonance in Fig. 4b (near 1535 nm), while in the undoped sample, no increase is observed. This shows that the increase in output power is due to the erbium emission, and is not reminiscent light from the pump. In both cases, there is a red-shift in the resonance, caused mainly by thermal effects [10]. By varying the pump power, we can measure the increase in output for the different resonances, as shown in Fig. 4c. Since it is not trivial to know exactly the propagation and coupling losses for the device, we cannot present the results as total gain. Instead, we plot the result as the ratio between the pumped and un-pumped output powers, in dB. This is also known as On/Off Gain, and can also be interpreted as loss reduction. It increases along with the pump power and no saturation can be observed, possibly due to the coupling losses reducing the effective pump and signal powers inside the ring. The gray curve in Fig. 4c is for the case where the signal is resonant within the internal rings. The On/Off gain is significantly higher than for the external ring resonances. However, as we can see from the spectra shown in Fig. 4a, even with the increase, the output power is small, limiting the applications of this resonance for the measured device design. This is mainly due to the non-optimized design for this particular case.

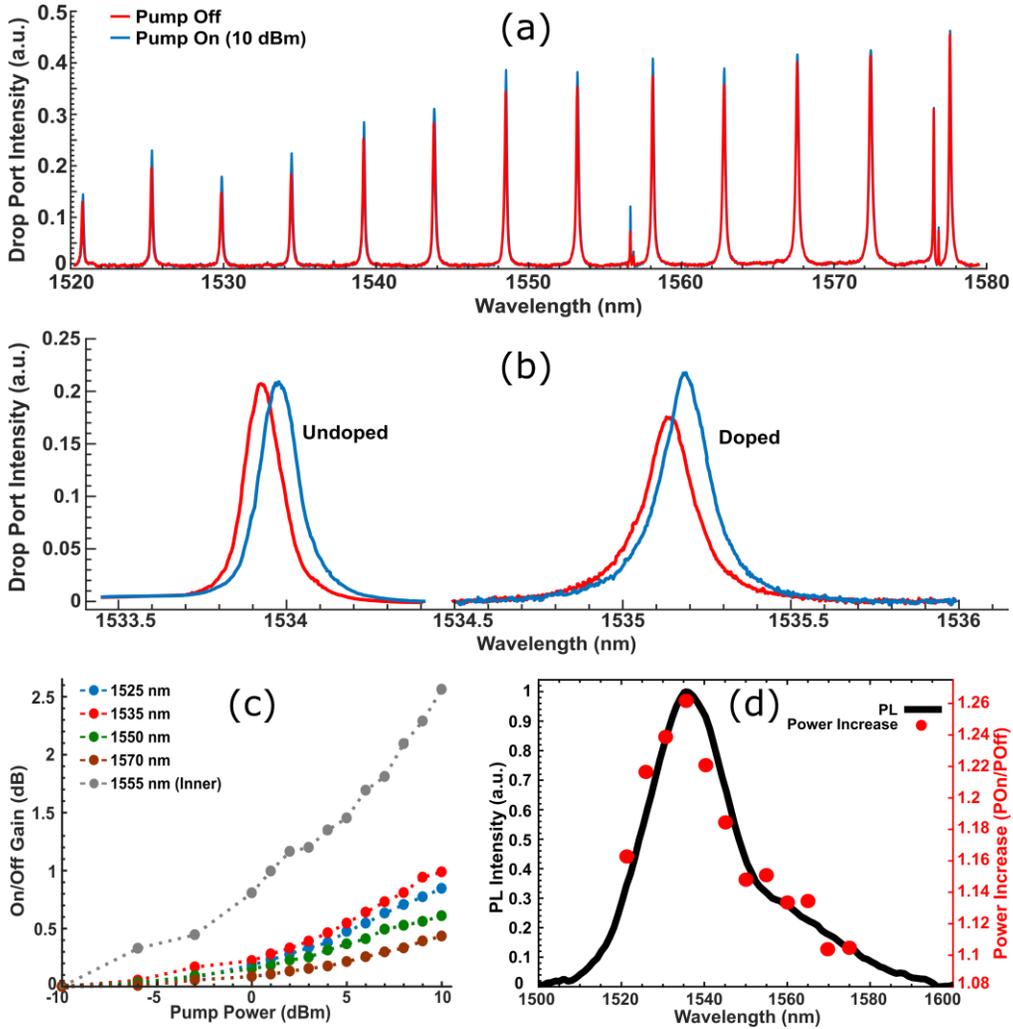

Fig. 4: Experimental results. (a) Broadband drop-port spectra for a doped device; (b) Comparison between undoped and doped samples. (c) On/Off gain vs. Pump Power for various resonances; (d) Photoluminescence measurement (solid black line) and output power increase for all resonances within the 1520-1580 nm range (red dots).

The power enhancement is spectrally inhomogeneous and reproduces the emission spectrum of the Er-doped cladding (Fig.4(d)). In this figure, the red dots are the power enhancement at maximum pump while the solid line shows the photoluminescence spectrum of the cladding film.

Comparing these results with those in Fig. 2, we observe that the predictions from our model are consistent with the experimental data. The measured increase in output power deviates only 10% from the calculated value, and the behavior when varying the pump power and resonance wavelength are equivalent for both the modelled and measured results. Thus, the model can be safely used to predict the results we could obtain from a device with different design parameters. As a first example, for a coupling gap that results in a Q Factor of 30000, we obtain a minimal insertion loss of 3 dB. However, when the cladding is considered to be Erbium-doped, there is a net gain of 1.8 dB instead (Fig. 5a). On the other hand, although the measured output power of the inner ring resonance is very small, with optimal design its efficiency can be increased, as the respective curve in Fig. 5b indicates, increasing the net gain to 4 dB.

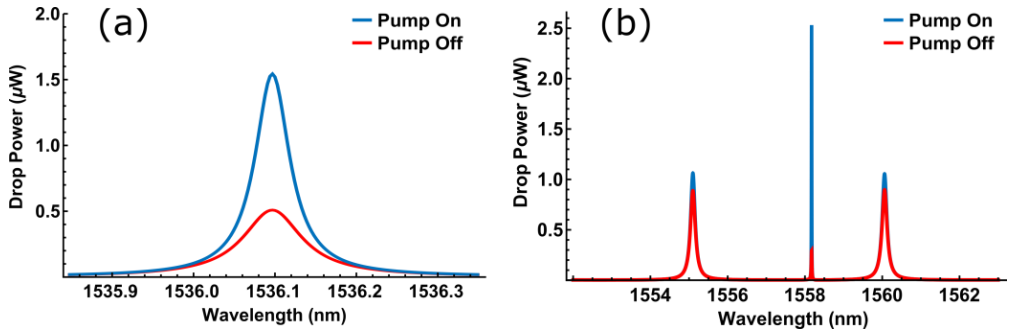

Fig. 5: (a) Prediction of a component that compensates the loss of a resonator with a Q Factor of 30000. (b) Using an optimized design for the resonances of an inner ring, a higher gain can be expected (4 dB on this case).

The downside of achieving considerable gain by increasing the Q factor is the decrease of the maximum bitrate supported. As the Q factor increases, there will be a situation where bits start interfering with one another, since there can be more than one bit confined in the cavity at the same time. Our measured device supports rates up to 40 Gb/s, while in the design discussed above the maximum bit rate decreases to 15 Gb/s. For most applications, it would still be suitable, however.

Additional applications of internally coupled rings arise when the resonances from both external and internal rings match. Instead of having one peak of transmission, corresponding to all the matching rings, there will be several, with a very short separation between them (< 0.5 nm). This is called resonance splitting, analogous to the splitting in atomic coupling levels. These split resonances may have a much higher Q (up to 3 times higher) than the isolated resonance. Thus, it is possible to design, by spectral engineering, complex filters that are appropriate for different specific applications [11], such as multiple channel routing. In the context of our component, with proper and careful design, it is possible to use this Q enhancement to achieve higher levels of amplification, or even design compact lasers with Er-doped claddings.

## 4.  Conclusion

We proposed and demonstrated a device that is very easily fabricated and can assist in overcoming insertion losses of silicon photonics filters and routers based on ring resonators. It is especially useful in complex routing circuits, where many rings are concatenated. Whereas regular amplifying methods employ EDWAs that can occupy precious chip real-estate, our proposal does the same without the need of sacrificing any additional area. As a final remark, this method of resonant amplifying has the potential to reach almost 2 dB of net gain in $2\times10^{-3}$ mm$^2$ (1000 dB/mm$^2$), while standard EDWAs can reach 20 dB of net gain in 60 mm$^2$ (0.3 dB/mm$^2$).

### Funding

São Paulo Research Foundation (FAPESP) grants 08/57857-2 and 2014/04748-2; National Council for Scientific and Technological Development (CNPq) grant 574017/2008-9.